\documentclass[manuscript]{aastex}

\begin{document}

\title{Response of bare strange stars to energy input onto their
surfaces}

\author{Vladimir V.~Usov}

\affil{Department of Condensed Matter Physics, Weizmann Institute,
Rehovot 76100, Israel}


\begin{abstract}

We study numerically the thermal emission of $e^+e^-$ pairs from a
bare strange star heated by energy input onto its surface; heating
starts at some moment, and is steady afterwards. The thermal
luminosity in $e^+e^-$ pairs increases to some constant value. The
rise time and the steady thermal luminosity are evaluated. Both
normal and colour superconducting states of strange quark matter
are considered. The results are used to test the magnetar model of
soft gamma-ray repeaters where the bursting activity is explained
by fast decay of superstrong magnetic fields and heating of the
strange star surface. It is shown that the rise times observed in
typical bursts may be explained in this model only if strange
quark matter is a superconductor with an energy gap of more that
1~MeV.


\end{abstract}


\section{Introduction}
Strange stars are astronomical compact objects which are entirely
made of deconfined quarks (for a review, see Glendenning 1996;
Weber 1999). The possible existence of strange stars is a direct
consequence of the conjecture that strange quark matter (SQM) may
be the absolute ground state of the strong interaction, i.e.,
absolutely stable with respect to $^{56}$Fe (Bodmer 1971; Witten
1984). SQM with a density of $\sim 5\times 10^{14}$ g~cm$^{-3}$
might exist up to the surface of a strange star. Recently, the
thermal emission from bare SQM surfaces of strange stars was
considered (Usov 1998, 2001a). It was shown that the surface
emissivity of SQM in both equilibrium photons and $e^+e^-$ pairs
created by the Coulomb barrier at the SQM surface is $\gtrsim
10$\% of the black body surface emissivity at the surface
temperature $T_{\rm s}\gtrsim 1.5\times 10^9$~K. Below this
temperature, $T_{\rm s} < 1.5\times 10^9$~K, the SQM surface
emissivity decreases rapidly with decrease of $T_{\rm s}$.

At the moment of formation of a strange star the surface
temperature may be as high as $\sim 10^{11}$~K (e.g., Haensel,
Paczy\'nski, \& Amsterdamski 1991). Since SQM at the surface of a
bare strange star is bound via strong interaction rather than
gravity, such a star can radiate at the luminosity greatly
exceeding the Eddington limit, up to $\sim 10^{52}$~ergs~s$^{-1}$
at $T_{\rm s}\sim 10^{11}$~K. (Alcock, Farhi, \& Olinto 1986;
Chmaj, Haensel, \& Slomi\'nski 1991; Usov 1998, 2001a). A young
strange star cools rapidly, and within about a month after its
formation the surface temperature is less than $2.5\times 10^8$~K
(e.g., Pizzochero 1991). In this case, the thermal luminosity from
the stellar surface in both equilibrium photons and $e^+e^-$ pairs
is negligibly small, $L_{\rm th}< 10^{25}$~ergs~s$^{-1}$. A bare
strange star with such a low surface temperature may be a strong
source of radiation only if its surface is reheated. Recently,
response of a bare strange star to accretion of a massive
comet-like object with the mass $\Delta M\sim 10^{25}$~g onto the
stellar surface was considered (Usov 2001b). It was shown that the
light curves of the two giant bursts observed from the soft
$\gamma$-ray repeaters SGR 0526-66 and SGR 1900+14 may be easily
explained in this model.

In this paper we report on our numerical simulations of the
response of a bare strange star to energy input onto its surface.
We consider a wide range of the rate of the energy input. Both
normal and colour superconducting SQM are discussed.

\section{Formulation of the problem}

The model to be studied is the following. The energy input onto
the surface of a bare strange star starts at the moment $t=0$, and
it is spherical and steady at $t\geq 0$. Since in our simulations
the surface temperature is not higher than ${\rm a~few}\times
10^9$~K, $e^+e^-$ pairs created by the Coulomb barrier are the
main component of the thermal emission from the stellar surface
(Usov 2001a). We assume that the process of the energy input has
no effect on the outflow of both created pairs and photons which
form due to annihilation of some of these pairs (cf. Usov 2001b).

The equation of heat transfer that describes
the temperature distribution at the surface layers of
a strange star is

\begin{equation}
C{\partial T\over \partial t}={\partial \over \partial x}
\left(K{\partial T\over \partial x}\right) -\varepsilon_\nu\,,
\end{equation}

\noindent
where $C$ is the specific heat for SQM per unit volume,
$K$ is the thermal conductivity, and $\varepsilon_\nu$
is the neutrino emissivity.

The heat flux due to thermal conductivity is

\begin{equation}
q=-K dT/dx\,.
\end{equation}

\noindent At the stellar surface, $x=0$, the heat flux directed
into the strange star is equal to (Usov 2001b)

\begin{equation}
q=L_{\rm input}/(4\pi R^2)-\varepsilon_\pm f_\pm\,,
\end{equation}

\noindent
where $L_{\rm input}$ is the rate of the energy input onto the
stellar surface, $R\simeq 10^6$~cm is the radius of the star,
$\varepsilon_\pm f_\pm$ is the energy flux in $e^+e^-$ pairs
emitted from the SQM surface, $\varepsilon_\pm\simeq m_ec^2 +
kT_{\rm s}$ is the mean energy of created pairs,

\begin{equation}
f_\pm\simeq 10^{39.2}\,T_{\rm s,9}
^3\exp \left(-{11.9\over
T_{\rm s,9}}\right)J(\zeta)\,\,\,{\rm cm}^{-2}\,{\rm s}^{-1}
\end{equation}

\noindent
is the flux of pairs from the unit SQM surface,

\begin{equation}
J(\zeta )={1\over 3}{\zeta^3\ln \, (1+2\zeta ^{-1})\over
(1+0.074\zeta )^3}
+ {\pi^5\over6}{\zeta^4\over (13.9 +\zeta)^4}\,,
\end{equation}

\noindent
$\zeta\simeq (2\times 10^{10}~{\rm K})/T_{\rm s}$, and
$T_{\rm s,9}$ is the surface temperature in units of $10^9$~K.

Eqs. $(2)-(5)$ give a boundary condition on $dT/dx$ at the stellar
surface.  We assume that at the initial moment,
$t=0$, the temperature in the surface layers is constant,
$T=10^8$~K.

It has been suggested (Bailin \& Love 1979, 1984) that the quarks
may eventually form Cooper pairs. Recently, superconductivity of
SQM was considered in detail (for a review, see Rajagopal, \&
Wilczek 2000; Alford, Bowers, \& Rajagopal 2001), and it was shown
that SQM is plausibly a colour superconductor if its temperature
not too high. Below, we consider both normal and superconducting
SQM.

\subsection{The Normal State of SQM}
For non-superconducting SQM, the contribution of the quarks to
both the specific heat and the thermal conductivity prevails over
the contributions of the electrons, photons and gluons. In this
case we have (Iwamoto 1982; Heiselberg \& Pethick 1993; Benvenuto
\& Althaus 1996)

\begin{equation}
C\simeq C_q\simeq 2.5\times 10^{20}(n_b/n_0)^{2/3}T_9\,\,\,
{\rm ergs~cm}^{-3}~{\rm K}^{-1}
\end{equation}

\begin{equation}
K\simeq K_q\simeq 6\times 10^{20}\alpha_c^{-1}
(n_b/n_0)^{2/3}\,\,\,
{\rm ergs}~{\rm cm}^{-1}~{\rm s}^{-1}~{\rm K}^{-1}
\end{equation}

\begin{equation}
\varepsilon_\nu \simeq 2.2\times 10^{26}\alpha_c
Y^{1/3}_e(n_b/n_0)T^6_9\,\,\,{\rm ergs}~{\rm cm}^{-3}~{\rm s}^{-1}
\end{equation}

\noindent
where
$n_0\simeq 1.7\times 10^{38}$~cm$^{-3}$ is normal nuclear matter
density, $n_b$ is the baryon number density of SQM,
$\alpha_c=g^2/4\pi$ is the QCD fine structure constant,
$g$ is the quark-gluon coupling constant,
$Y_e=n_e/n_b$ is the number of electrons per baryon,
and $T_9$ is the temperature in units of $10^9$~K.

\subsection{The Colour Superconducting State of SQM}
SQM may be a colour superconductor if its temperature is below
some critical value. In the classic model of Bardeen, Cooper, and
Schrieffer the critical temperature is $T_c\simeq
0.57\Delta_0/k_{_{\rm B}}$, where $\Delta_0$ is the energy gap at
zero temperature and $k_{_{\rm B}}$ is the Boltzmann constant
(e.g., Carter \& Reddy 2000). The value of $\Delta _0$ is very
uncertain and lies in the range from $\sim 0.1-1$~MeV (Bailin \&
Love 1984) to $\sim 50-10^2$~MeV (Alford, Rajagopal, \& Wilczek
1998; Alford, Berges, \& Rajagopal 1999; Pisarski \& Rischke
2000).

We use the following interpolation formula for the temperature
dependence of the energy gap(e.g., Carter \& Reddy 2000):

\begin{equation}
\Delta (T)= \Delta _0\left[1-\left({T\over T_c}\right)^2
\right]^{1/2}.
\end{equation}

Superconductivity modifies the properties of SQM significantly.
The specific heat of superconducting SQM increases discontinuously
as the temperature falls below the critical temperature, and then
decreases exponentially at lower temperatures. For the quark
specific heat at $T\leq T_c$ we adopt the theoretical results of
M\" uhlschlegel (1959) for superconducting nucleons, i.e.,
(Horvath, Benvenuto, \& Vucetich 1991; Blaschke, Kl\" ahn, \&
Voskresensky 2000)

$$
\tilde C_q=3.2 C_q ({T_c/ T})
\,\exp\, (-{\Delta_0/ T})
$$

\begin{equation}
\times [2.5 -1.7T/ T_c+3.6({T/ T_c})^2]\,,
\end{equation}

\noindent here and below tilde signifies that this value relates
to superconducting SQM.  Even at $T\ll T_c$ the suppression of the
specific heat of SQM is never complete because the electrons
remain unpaired. The specific heat of superconducting SQM which is
used in our simulations is $\tilde C\simeq \tilde C_q+C_e$, where

\begin{equation}
C_e\simeq 5.7\times 10^{19}Y_e^{2/3}(n_b/n_0)^{2/3}T_9
\,\,\,{\rm ergs~cm}^{-3}~{\rm K}^{-1}\,,
\end{equation}

\noindent is the specific heat of the electron subsystem of SQM
(Blaschke, Grigorian, \& Voskresensky 2001).

At $T < T_c$ both the thermal conductivity of SQM and its neutrino
emissivity are suppressed by a factor of
$\exp\,(-\Delta/k_{_{\rm B}}T$). In our simulations, we adopt

\begin{equation}
\tilde K = K_q \exp [-\Delta (T)/k_{_{\rm B}}T]+K_e\,,
\end{equation}

\begin{equation}
\tilde \varepsilon_\nu = \varepsilon_\nu \exp [-\Delta (T)
/k_{_{\rm B}}T]\,,
\end{equation}

\noindent
where $K_q$ and $\varepsilon_\nu$ are given by equations (7)
and (8), respectively, and

\begin{equation}
K_e\simeq 5.5\times 10^{23}Y_e(n_b/n_0)T_9^{-1}\,\,\,
{\rm ergs~cm}^{-1}~{\rm s}^{-1}~{\rm K}^{-1}
\end{equation}

\noindent is the thermal conductivity of the electrons (e.g.,
Blaschke et~al. 2001)

\section{Results of numerical simulations}
The set of equations $(1)-(5)$ was solved numerically for both
normal $(\Delta_0=0)$ and superconducting $(\Delta_0 > 0)$ states of
SQM. We assumed the typical values of $\alpha_c=0.1$, $n_b=2n_0$,
and $Y_e=10^{-4}$. Figure~1 shows a typical temporal behaviour of
the strange star luminosity, $L_\pm = 4\pi R^2\varepsilon_\pm f_\pm$,
in $e^+e^-$ pairs. From this Figure, we can see that
$L_\pm$ increases eventually to its maximum value $L_\pm^{\rm max}$.
The rate of this increase may be characterized by
the rise time $\tau_\pm$ that is determined as a time interval from
the initial moment, $t=0$, to the moment when $L_\pm$ is equal to
$(1/2)L_\pm^{\rm max}$. The results of our simulations are presented
in Tables~1 and~2. There is a critical
value, $L_{\rm cr}$, of the input luminosity at which the dependence
of $L_\pm^{\rm max}$ on $L_{\rm input}$ changes qualitatively
(see Table~1). $L_{\rm cr}$ is $\sim 10^{40}$~ergs~s$^{-1}$ for
normal SQM and $\sim 3\times 10^{38}$~ergs~s$^{-1}$
for superconducting SQM with $\Delta_0\gtrsim 0.3$~MeV. At $L_
{\rm input} \gtrsim {\rm a~few}\times L_{\rm cr}$, $L_\pm^{\rm max}$
is about $L_{\rm input}$ while when $L_{\rm input}$ is a few times below
than $L_{\rm cr}$ the thermal emission of $e^+e^-$ pairs from
the stellar surface is negligible.

In our simulations, the rise time $\tau_\pm$ varies in a
very wide range from $\sim 10^6$~s
at $L_{\rm input}\lesssim 10^{40}$~ergs~s$^{-1}$ and $\Delta_0\lesssim
0.1$~MeV to $\sim 10^{-8}$~s at $L_{\rm input}\sim
10^{45}$~ergs~s$^{-1}$ and $\Delta_0\gtrsim 1$~MeV (see Table~2).

At $\Delta_0\gtrsim 1$~MeV the thermal emission of $e^+e^-$ pairs
does not depend on $\Delta_0$ because in this case both the specific
heat of the quarks and their thermal conductivity are strongly
suppressed, and the heat transport is mostly determined by
the electron subsystem of SQM.

The rise time of the luminosity in neutrinos is many orders of
magnitude larger that $\tau_\pm$, especially  when $\tau_\pm$ is
small. In our model the neutrino luminosity may increase up to
$L_{\rm input}-L_\pm^{\rm max}$ when $t$ goes to infinity. This is
because all energy which is delivered onto the stellar surface is
radiated either from the surface by $e^+e^-$ pairs or from the
stellar interior by neutrinos.

\section{Discussion}
Since bare strange stars can radiate at the luminosities greatly
exceeding the Eddington limit (Alcock et al. 1986; Chmaj et al.
1991; Usov 1998, 2001a), these stars are reasonable candidates for
soft $\gamma$-ray repeaters (SGRs) that are the sources of brief
($\sim 10^{-2}-10^2$~s) bursts with Super-Eddington luminosities,
up to $\sim 10^{43}-10^{45}~{\rm ergs~s}^{-1}$. The bursting
activity of a SGR may be explained by fast heating of the bare,
rather cold ($T_{\rm s}\lesssim 10^8$~K) surface of a strange star
up to the temperature of $\sim (1-2)\times 10^9$~K and its
subsequent thermal emission (Usov 2001a). The heating mechanism
may be either fast decay of superstrong ($\sim 10^{14}-10^{15}$~G)
magnetic fields (Usov 1984; Duncan \& Thompson 1992; Paczy\' nski
1992; Thompson \& Duncan 1995; Cheng \& Dai 1998; Heyl \& Kulkarni
1998) or impacts of comet-like objects onto the stellar surface
(Harwitt \& Salpeter 1973; Newman \& Cox 1980; Zhang, Xu, \& Qiao
2000; Usov 2001b). The magnetar model of SGRs which is based on
the first mechanism is most popular now. In this model, the
magnetic energy of a strongly magnetized strange star (magnetar)
may be released from time to time due to MHD instabilities. A
violent release of energy inside a magnetar excites surface
oscillations (e.g., Thompson \& Duncan 1995). In turn, this
shaking may generate strong electric fields in the magnetar
magnetosphere which accelerate particles to high energies. These
high energy particles bombard the surface of the strange star and
heat it. At the input luminosities of $\sim
10^{41}-10^{45}$~ergs~s$^{-1}$, which are typical for SGRs, the
efficiency of reradiation of the partical energy by the stellar
surface to $e^+e^-$ pairs is very high, $L_\pm/L_{\rm input}\simeq
1$, especially for superconducting SQM (see Table~1). Since for
SGRs the burst luminosities are at least a few orders of magnitude
higher than $L_*\simeq 4\pi m_{\rm e}c^3R/\sigma_{_{\rm T}}\simeq
10^{36}~{\rm ergs~s}^{-1}$, the outflowing $e^+e^-$ pairs mostly
annihilate in the vicinity of the strange star (e.g., Beloborodov
1999). Therefore, at $L_{\rm input}\sim
10^{41}-10^{45}$~ergs~s$^{-1}$ far from the star the luminosity in
X-ray and $\gamma$-ray photons practically coincides with $L_{\rm
input}$, $L_\gamma\simeq L_\pm -L_*\simeq \L_\pm \simeq L_{\rm
input}$.

Two giant bursts were observed on 5 March 1979 and 27 August 1998
from SGR~$0526-66$ and SGR~$1900+14$, respectively. The peak
luminosities of these bursts were $\sim 10^{45}$~ergs~s$^{-1}$
(Fenimore, Klebesadel, \& Laros 1996; Hurley et al. 1999). In this
case, from Table~2 the rise time expected in our model is
$\lesssim 10^{-3}$~s that is consistent with available data on the
two giant bursts (Mazets et al. 1999). This is valid irrespective
of that SQM is a colour superconductor or not. For typical bursts
of SGRs the luminosities are $\sim 10^{41}-10^{42}$~ergs~s$^{-1}$
(e.g., Kouveliotou 1995), and the observed rise times ($\sim
10^{-1}-10^{-3}$~s) may be explained in our model only if SQM is a
superconductor with the energy gap $\Delta_0 \gtrsim 1$~MeV (see
Table~2).

\begin{acknowledgements}
This work was supported by the Israel Science Foundation of
the Israel Academy of Sciences and Humanities.
\end{acknowledgements}

\onecolumn

\clearpage

\begin{table}
\caption{The maximum luminosity $L_\pm^{\rm max}$ in $e^+e^-$ pairs}
$$
\begin{array}{rrrrrr}
\tableline
\hline
\noalign{\smallskip} 
\,\,L_{\rm input},\, &  \multicolumn{5}{c}{ {\rm The~energy~gap}~\Delta\,,~\,
{\rm MeV}}    \\
\cline{2-6} 
\,\,{\rm ergs~s}^{-1}\, &0\,\,\, &0.1\,\,\, & 0.3\,\,\, & 1\,\,\, & 10 \,\,\,\\
\noalign{\vskip5pt\hrule\vskip3pt}
10^{38}     & < 10^{15} &  < 10^{15} & \sim 10^{25}   & \sim 10^{25}
&\sim 10^{25} \\
3\times 10^{38}& <10^{15}& 3\times 10^{33} & 9.6\times 10^{37} &
9.9\times 10^{37} &  10^{38} \\
10^{39}  & \sim 10^{17}  &  2\times 10^{36}  & 6.8\times 10^{38} &
7.2\times 10^{38}  &  7.3\times 10^{38}  \\
3\times 10^{39}   & 10^{31} &  8\times 10^{38}  & 2.67\times 10^{39} &
2.72\times 10^{39}  &   2.72\times 10^{39}   \\
10^{40} & 1.1\times 10^{39}  & 4.7\times 10^{39}  & 9.6\times 10^{39} &
9.7\times 10^{39}  &  9.7\times 10^{39} \\
10^{41}  & 8.4\times 10^{40} & 8.5\times 10^{40}  &  9.93\times 10^{40}
& 9.96\times 10^{40} &  9.96\times 10^{40} \\
10^{42}   & 9.76\times 10^{41} & 9.79\times 10^{41} & 9.991\times 10^{41} &
\,\,\,\,\,9.996\times 10^{41}  & \,\,\,\,\,9.996\times 10^{41} \\
10^{43} & 9.96\times 10^{42} & 9.96\times 10^{42} & 9.998\times 10^{42} &
10^{43} & 10^{43} \\
10^{44} & 9.992\times 10^{43} & 9.992\times 10^{43}
& \,\,\,\,\,9.999\times 10^{43} & 10^{44}   & 10^{44}   \\
10^{45}   & \,\,\,\,\,9.999\times 10^{44} & \,\,\,\,\,9.999\times 10^{44}
& 10^{45}  &
10^{45}  & 10^{45} \\
\end{array}
$$

\tablecomments{Units of $L_\pm^{\rm max}$ are ergs~s$^{-1}$.
In all cases when in this Table $L_\pm^{\rm max}$ is equal to
$L_{\rm input}$ we have $(L_{\rm input} -L_\pm^{\rm max})/ L_{\rm input}
< 10^{-4}$.}

\end{table}

\clearpage

\begin{table}
\caption{The rise time $\tau_\pm$ for the thermal emission of
$e^+e^-$ pairs}
$$
\begin{array}{rrrrrr}
\tableline
\hline
\noalign{\smallskip} 
\,\,L_{\rm input},\, &  \multicolumn{5}{c}{ {\rm The~energy~gap}~\Delta\,,~\,
{\rm MeV}}    \\
\cline{2-6} 
\,\,{\rm ergs~s}^{-1}\, &  0\,\,\,& 0.1\,\,\, & 0.3\,\,\,& 1\,\,\,& 10 \,\,\,\\
\noalign{\vskip5pt\hrule\vskip3pt}
10^{38} & \sim 10^6 & \sim 10^6  & 10^4 &  10^4 &10^4  \\
3\times 10^{38}  & \sim 10^6  &  1.2\times 10^6  &  2\times 10^4 &
6\times 10^3  &6\times 10^3  \\
10^{39}  &  6\times 10^5  &  10^6  &3.6\times 10^3  &
7.6\times 10^2  &  7.7\times 10^2  \\
3\times 10^{39}  & 9\times 10^5  &  6\times 10^5  &  9\times 10^2  &
1.2\times 10^2 &   1.2 \times 10^2  \\
10^{40}  &  4.4\times 10^5  &3.4\times 10^4  &1.5\times 10^2  &
13.5  &  13.5  \\
10^{41}  &  1.3\times 10^4  &   3.5 \times 10^3  & 4.8 & 0.19  & 0.19 \\
10^{42}  & 2.4\times 10^2  & 1.5\times 10^2 & 0.18 & 2.5\times 10^{-3}  &
2.5\times 10^{-3}  \\
10^{43} & 3.9 &2.5 & 9\times 10^{-3}  & 3.6\times 10^{-5}  &
3.6\times 10^{-5}  \\
10^{44} & 6.7\times 10^{-2}  & 5.6\times 10^{-2}  & 5.5\times 10^{-4} &
5.6\times 10^{-7} & 5.6\times 10^{-7}  \\
10^{45} & \,\,\,\,\,\,\,\,\,1.3\times 10^{-3}&\,\,\,\,\,\,\,\,\,1.2\times
10^{-3} & \,\,\,\,\,\,\,\,\,4.6\times 10^{-5} &\,\,\,\,\,\,\,\,\,
1.3\times 10^{-8} & \,\,\,\,\,\,\,\,\,0.8\times 10^{-8}  \\
\end{array}
$$

\tablecomments{Units of $\tau_\pm$ are seconds.}

\end{table}

\clearpage


\begin{figure}
\plotone{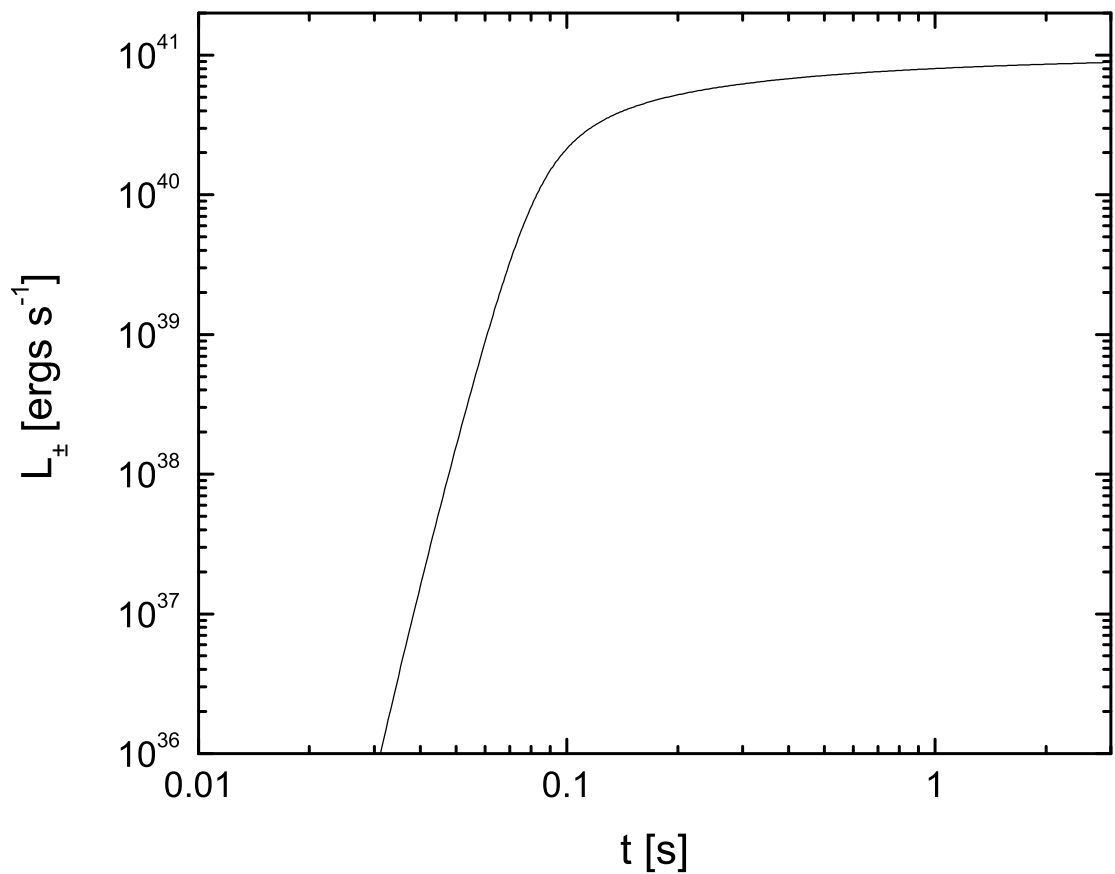}
\caption{The thermal luminosity in $e^+e^-$
pairs as a function of time for $L_{\rm
input}=10^{41}$~ergs~s$^{-1}$ and $\Delta_0=1$~MeV. \label{fig1}}
\end{figure}

\end{document}